# LEIA Discovery of the Longest-lasting and Most Energetic Stellar X-Ray Flare Ever Detected

Xuan Mao[1,2], He-Yang Liu[1], Song Wang[1,3], Zhixing Ling[1,2,3], Weimin Yuan[1,2], Huaqing Cheng[1], Haiwu Pan[1], Dongyue Li[1], Fabio Favata[4,5], Tuo Ji[6], Jujia Zhang[7,8,9], Xinlin Zhao[1,2], Jing Wang[1,10,11], Mingjun Liu[1,2], Yuan Liu[1], Zhiming Cai[12], Alberto J. Castro-Tirado[13,14], Yanfeng Dai[1], Licai Deng[1,2,15], Xu Ding[7,9,16], Kaifan Ji[2,7,9,16], Chichuan Jin[1,2,3], Yajuan Lei[1], Huali Li[1], Jun Lin[2,7,16,17], Huaqiu Liu[12], Shuai Liu[1], Hui Sun[1], Shengli Sun[18], Xiaojin Sun[18], Jianrong Shi[1,2], Jianguo Wang[7,9], Jingxiu Wang[1,2], Wenxin Wang[1], Jianyan Wei[1,2], Liping Xin[1], Dingrong Xiong[7], Chen Zhang[1,2], Wenda Zhang[1], Yonghe Zhang[12], Xiaofeng Zhang[12], Donghua Zhao[1], and Guiping Zhou[1,2]

[1] National Astronomical Observatories, Chinese Academy of Sciences, Beijing 100101, People's Republic of China; liuheyang@nao.cas.cn, songw@bao.ac.cn, lingzhixing@bao.ac.cn, wmy@nao.cas.cn
[2] School of Astronomy and Space Science, University of Chinese Academy of Sciences, Chinese Academy of Sciences, Beijing 100049, People's Republic of China
[3] Institute for Frontiers in Astronomy and Astrophysics, Beijing Normal University, Beijing 102206, People's Republic of China
[4] INAF—Osservatorio Astronomico di Palermo, Piazza del Parlamento, 1, 90134 Palermo, Italy
[5] Department of Physics, Imperial College London, Exhibition Road, London SW7 2AZ, UK
[6] Polar Research Institute of China, Shanghai 200136, People's Republic of China
[7] Yunnan Observatories, Chinese Academy of Sciences, Kunming 650216, People's Republic of China
[8] International Centre of Supernovae, Yunnan Key Laboratory, Kunming 650216, People's Republic of China
[9] Key Laboratory for the Structure and Evolution of Celestial Objects, Chinese Academy of Sciences, Kunming 650216, People's Republic of China
[10] Guangxi Key Laboratory for Relativistic Astrophysics, School of Physical Science and Technology, Guangxi University, Nanning 530004, People's Republic of China
[11] GXU-NAOC Center for Astrophysics and Space Sciences, Nanning 530004, People's Republic of China
[12] Innovation Academy for Microsatellites, Chinese Academy of Sciences, Shanghai 201210, People's Republic of China
[13] Instituto de Astrofísica de Andalucía (IAA-CSIC), Granada, Spain
[14] Unidad Asociada al CSIC, Departamento de Ingeniería de Sistemas y Automática, Escuela de Ingenierías, Universidad de Málaga, Málaga, Spain
[15] Department of Astronomy, China West Normal University, Nanchong 637002, People's Republic of China
[16] Center for Astronomical Mega-Science, Chinese Academy of Sciences, Beijing 100012, People's Republic of China
[17] Yunnan Key Laboratory of Solar Physics and Space Science, Kunming 650216, People's Republic of China
[18] Shanghai Institute of Technical Physics, Chinese Academy of Sciences, Shanghai 200083, People's Republic of China

## Abstract

The Lobster Eye Imager for Astronomy (LEIA) detected a new X-ray transient on 2022 November 7, identified as a superflare event occurring on a nearby K-type giant star HD 251108. The flux increase was also detected in follow-up observations at X-ray, UV, and optical wavelengths. The flare lasted for about 40 days in soft X-ray observations, reaching a peak luminosity of $\sim 1.1 \times 10^{34}$ erg s$^{-1}$ in 0.5–4.0 keV, which is roughly 60 times the quiescent luminosity. Optical brightening was observed for only one night. The X-ray light curve is well described by a double fast rise and exponential decay model, attributed to the cooling process of a loop arcade structure formed subsequent to the initial large loop with a half-length of $\sim 1.9 \times 10^{12}$ cm. Time-resolved X-ray spectra were fitted by a four-temperature apec model (with three components being the quiescent background), showing significant evolution of plasma temperature and emission measure over time. The estimated energy released in the LEIA band is $\sim 3 \times 10^{39}$ erg, suggesting that this is likely the most energetic X-ray stellar flare with the longest duration detected to date.

*Unified Astronomy Thesaurus concepts:* Stellar activity (1580); X-ray transient sources (1852); Stellar flares (1603)

## 1. Introduction

Solar and stellar flares are sudden and intense electromagnetic radiation enhancement detectable across a wide range of frequencies typically lasting from minutes to hours. These flares are believed to be caused by the rapid release of magnetic energy during the impulsive reconnection of twisted magnetic fields in the outer atmosphere (e.g., K. Shibata & T. Magara 2011; L. M. Walkowicz et al. 2011). Although stellar flares are believed to be generated by similar processes as solar flares (see also J. Lin & T. G. Forbes 2000), the diverse characteristics of stellar species and their space environments can lead to a wider range of flare parameters, including peak luminosity, duration, and total energy release. In particular, some stellar flares, known as superflares, can release over 10 to a million times the energy of the largest solar flares ($\sim 10^{32}$ erg; A. G. Emslie et al. 2012) and have been observed in various stars (e.g., H. Maehara et al. 2012; T. Shibayama et al. 2013; S. Candelaresi et al. 2014; S. L. Hawley et al. 2014; J. R. A. Davenport 2016; Y. Notsu et al. 2019; S. Okamoto et al. 2021). Such superflares with extremely powerful bursts of energy released by stars can help give insight into both the nature of stellar activities and the potential challenges faced by planets in close proximity to them.

RS Canum Venaticorum (RS CVn) type binaries are composed of two late-type stars, typically with one being a giant or subgiant and the other a dwarf or subgiant. These binary systems are notable for their intense stellar activities, including highly energetic and prolonged flares (J. C. Pandey & K. P. Singh 2012; C. I. Martìnez et al. 2022) and large star spots due to magnetic interactions between the stars (A. J. Drake 2006; A. J. Drake et al. 2014). Violent flares lasting up to several days with released energies of $\sim 10^{38}$ erg have been reported on RS CVn-type Star GT Mus (R. Sasaki et al. 2021). Moreover, these systems generally have synchronized rotational and orbital periods due to strong tidal forces (S. Karmakar et al. 2023). The synchronization will enhance magnetic interactions and can lead to heightened stellar activity. These characteristics make RS CVn binaries valuable for studying stellar magnetism and dynamics in binary systems.

**Table 1**
Multiwavelength Observations on HD 251108

| Instrument | Obs-Date | Wave Band |
| --- | --- | --- |
| LEIA | from 2022-11-07 to 2022-11-18 | Soft X-ray |
| Swift | from 2022-11-09 to 2022-12-25 | Soft X-ray/UV |
| NICER | from 2022-11-09 to 2023-02-06 | Soft X-ray |
| ASAS-SN | from 2022-11-02 to 2023-02-09 | optical |
| GWAC | from 2022-11-07 to 2023-01-19 | optical |
| BOOTES-4/MET | from 2022-11-09 to 2022-11-22 | optical |
| Lijiang 2.4 m Telescope | 2022-11-10, 2022-11-30 and 2022-12-24 | optical |

HD 251108 is a nearby K2 type giant star located at R.A. = 06:04:15.0, decl. = 12:45:51, with a distance of $504.7^{+4.8}_{-4.6}$ pc (Gaia EDR3 measurement; C. A. L. Bailer-Jones et al. 2021). In a survey conducted by M. Kiraga (2012) using All Sky Automated Survey (ASAS) photometric data, HD 251108 was classified as a candidate RS CVn star; however, its binary nature remains uncertain. Using Gaia's data, F. Anders et al. (2019) estimated its effective temperature to be $T_{\mathrm{eff}} = 4545^{+405}_{-138}$ K, with surface gravity $\log g = 2.13^{+0.17}_{-0.09}$ [cgs] and a mass of $M = 1.14^{+0.48}_{-0.19} M_\odot$. HD 251108 is also identified as the X-ray source 2RXS J060415.1+124554, which is detected by ROSAT with a flux of $8.246 \times 10^{-12}$ erg s$^{-1}$ cm$^{-2}$ in 0.1–2.4 keV (T. Boller et al. 2016) and by eROSITA with a flux of $7.518 \times 10^{-12}$ erg s$^{-1}$ cm$^{-2}$ in 0.2–2.3 keV (P. Predehl et al. 2021; A. Merloni et al. 2024). The corresponding quiescent X-ray luminosity is $\sim 10^{32}$ erg s$^{-1}$.

The Lobster Eye Imager for Astronomy (LEIA, C. Zhang et al. 2022; Z. X. Ling et al. 2023) is an operative wide-field X-ray monitor (see Section 2.1 for a brief introduction). On 2022 November 7, a burst event was first detected by LEIA as a new X-ray transient, which was later identified as a superflare occurring on the star HD 251108. Following this detection, a multiwavelength campaign was conducted on this source, including observations and monitoring by Swift, Neutron Star Interior Composition Explorer (NICER), All Sky Automated Survey for SuperNovae (ASAS-SN), Ground-based Wide Angle Camera (GWAC), Burst Observer and Optical Transient Exploring System (BOOTES), and the Lijiang 2.4 m telescope (see Table 1). These observations revealed that the flare from HD 251108 was extraordinarily luminous and prolonged. The estimated energy release and duration suggests it may be the most energetic and longest-lasting X-ray stellar flare ever recorded.

In this work, we present our study of the exceptional stellar flare that occurred on HD 251108.[19] This paper is organized as follows. First, we describe the observation and data reduction of this source in X-ray, UV, and optical bands in Section 2. Next, we present a detailed analysis of time-resolved spectra and light curves in Section 3. Then, the physical scenarios of the flare cooling process and the stellar activity are discussed in Section 4, followed by a summary in Section 5.

## 2. Observations and Data Reduction

On 2022 November 7, LEIA detected a new X-ray transient, designated LXT 221107A (Z. X. Ling et al. 2022). Within the approximate 3′ error circle of this transient, there exists a known ROSAT source, 2RXS J060415.1+124554, associated with HD 251108. However, the flux of this known ROSAT source is 10 times lower than that of LXT 221107A. An optical counterpart coincided with HD 251108 was identified by GWAC on 2022 November 7. To explore the nature of LXT 221107A, we performed a Swift target of opportunity observation on 2022 November 9. The X-ray Telescope (XRT) on board Swift detected an X-ray source spatially consistent with HD 251108, and no additional candidates were found in the LXT 221107A localization error region. The XRT spectrum was well fitted by a collisionally ionized plasma model, giving a flux that was still significantly higher than the quiescent level of HD 251108. This observational evidence confirms that LXT 221107A was a superflare event on HD 251108. To study this exceptional flare in greater detail, further observations and monitoring over the following three months were conducted, as described in the subsequent sections.

### 2.1. LEIA

LEIA is a focusing X-ray telescope with a large spontaneous field-of-view (FoV) of about 340 square degrees and a soft X-ray energy bandpass of 0.5–4.0 keV, enabled by novel lobster eye micropore optics technology (Z. X. Ling et al. 2023). As a pathfinder of the Einstein Probe (EP) mission (W. Yuan et al. 2022), LEIA was launched on 2022 July 27 into a Sun-synchronous orbit with a height of 550 km on board the SATech-01 satellite of the Chinese Academy of Sciences (CAS). It has been operational since achieving its first light in 2022 August (C. Zhang et al. 2022). LEIA has a spatial resolution of $\sim 5'$ (full width at half maximum, FWHM) and a sensitivity of $2-3 \times 10^{-11}$ erg s$^{-1}$ cm$^{-2}$ with a 1000 s exposure.

LEIA detected LXT 221107A on 2022 November 7, and subsequently conducted a series of 40 observations to monitor the source from November 7 to November 18. During this period, LEIA found that the flux of LXT 221107A increased continuously, reaching its peak on 2022 November 8, before beginning a gradual decline.

---

[19] We note that an independent analysis was also performed by H. M. Günther et al. (2024) of the X-ray data from the NICER and Swift/XRT follow-up observations on this flare, which were triggered following the discovery by LEIA and are part of our work presented here. While the results of the two studies agree largely with each other, our work presents new data reporting the rising and peak phase of the flare detected by LEIA; we also find a prolonged flare duration than that in H. M. Günther et al. (2024), as well as a secondary flare much smaller than the primary one by using more observations taken with NICER during the later phase of the flare all the way to the quiescent state.

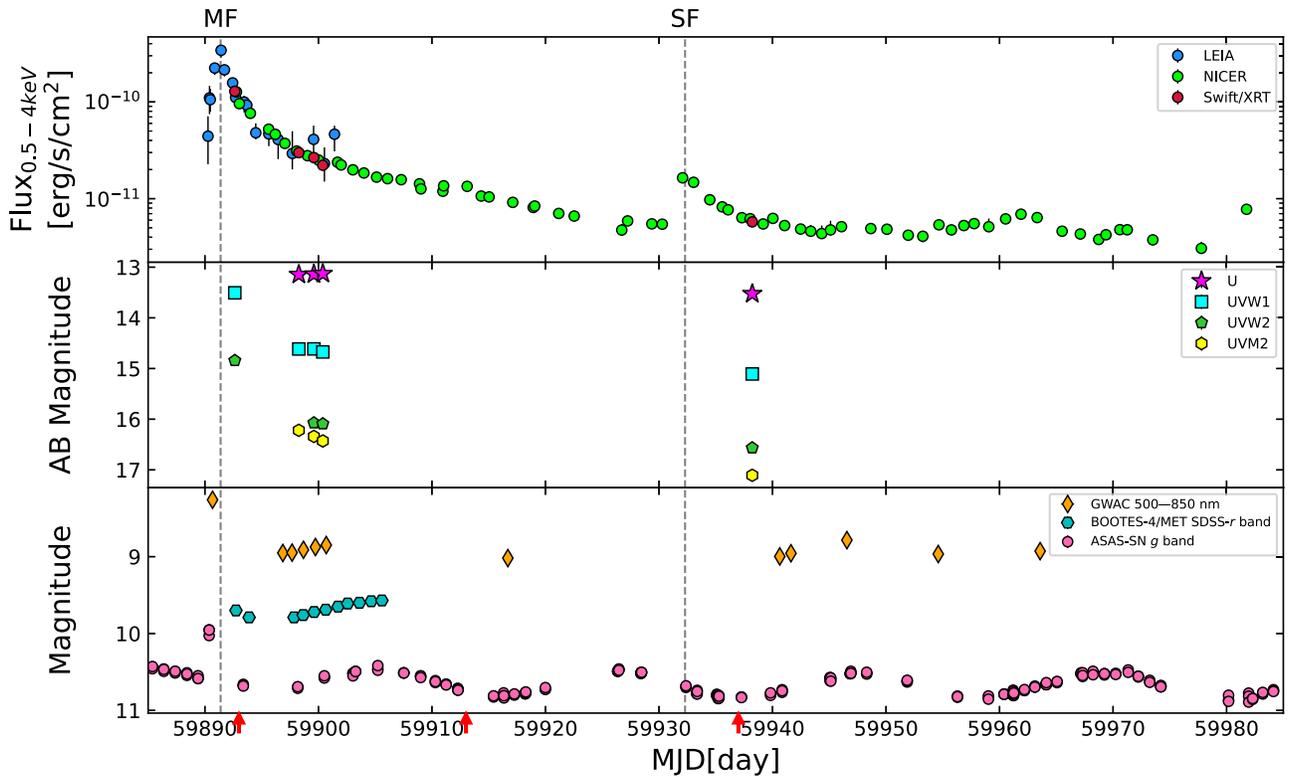

**Figure 1.** The comprehensive light curve of HD 251108 in soft X-ray (top panel), UVOT bands (middle panel), and ground-based optical photometry (bottom panel). For the soft X-ray (top panel), the unabsorbed flux in the 0.5–4.0 keV for each spectrum was estimated using the cflux model during spectral fitting in xspec (see Section 3.1 for more details on the spectral fitting process). The gray dashed lines mark the peak times of the main flare and the secondary flare. The red arrows at the bottom indicate when the three optical spectra were taken (see Section 2.8).

The LEIA data was reduced using the data reduction software developed for the EP mission (Y. Liu et al. 2025, in preparation), and the calibration database (CALDB) generated from both on-ground and in-orbit calibration campaigns (H. Q. Cheng et al. 2025, in preparation; H. Q. Cheng et al. 2024).

### 2.2. Swift/XRT

A Swift target of opportunity observation (Target ID: 15410, PI: D.Y. Li) was conducted following the detection of LXT 221107A by LEIA. The XRT onboard Swift began observations on 2022 November 9, and detected the source (D. Y. Li et al. 2022), noting that the flux was already lower than the peak level detected by LEIA. Over the following two months, we conducted four additional observations using Swift. The data of these five observations were downloaded from HEASARC, and were calibrated and screened through the standard reduction procedures using the HEAsoft package (v6.32.1) and Swift/XRT CALDB (version 20230725).

### 2.3. NICER

The first NICER observation of LXT 221107A began on 2022 November 9 (D. Pasham et al. 2022), and the observed flux was consistent with Swift/XRT's measurements (see Figure 1). NICER continued to observe the source for nearly three months, with observation IDs ranging from 5203530101 to 5203530169, totaling an exposure time of about 170 ks. These observations revealed that this event was an extremely long-lasting flare and detected another weaker flare about 40 days after LXT 221107A.

Data from all 69 NICER observations were obtained from the HEASARC website, except for observation ID 5203530129, which had an exposure time of 0 s. The data were processed using the HEAsoft package (v6.32.1), incorporating NICERDAS (version 10) and NICER/XTI CALDB (version 20221001).

### 2.4. Swift/UVOT

Along with XRT, the Ultraviolet/Optical Telescope (UVOT) onboard Swift also conducted observations of LXT 221107A. For each of five Swift observations, four UVOT filters—U, uvw1, uvw2, and uvm2—were used to capture the images, except during the second observation. Aperture photometry was performed using the uvotsource task with Swift/UVOTA CALDB (version 20240201). The U and uvm2 data from the first observation were excluded because the source was located in a bad area of the image.

### 2.5. ASAS-SN

We searched the ASAS-SN Sky Patrol v1.0 and examined the light curve of HD 251108, which was generated using Aperture Photometry. A brightening was observed in three consecutive observations in g band on the night of 2022 November 7, confirming that LXT 221107A was a high-level stellar flare occurring on HD 251108.

The bottom panel of Figure 1 shows all ASAS-SN g-band observations of HD 251108 during the flare period. The brightening was found only on the detection day of LXT 221107A, indicating that its optical flux returned to quiescent levels within three days. This return to quiescence is considerably quicker than that in the soft X-ray band. Such differing timing behaviors across wavelengths are common in solar and stellar flares (A. Benz 2002; A. O. Benz & M. Güdel 2010).

### 2.6. GWAC

During the X-ray flare, an optical brightening of about 0.6 mag relative to its quiescent level was also detected by GWAC (J. Wang et al. 2021; L. P. Xin et al. 2024) in its routine survey. As one of main ground facilities of the Space-based multiband Variable Objects Monitor (SVOM; J. Wei et al. 2016), GWAC monitored about 2000 square degrees of the sky in a cadence of 15 s. The observation was carried out in white filter, which was calibrated into the *R* band of the Johnson–Bessel system. The general behavior of the GWAC light curve was consistent with the observations of ASAS-SN. Each data point in Figure 1 represents the mean magnitude of the observations at that night.

### 2.7. BOOTES-4/MET

Follow-up optical observations were conducted using the BOOTES-4/MET 0.6 m optical telescope from November 9 to November 22 (D. R. Xiong et al. 2022). After applying flat field and bias corrections, aperture photometry was performed using PyRAF.[20] Two comparison stars, TYC 725-825-1 and TYC 725-489-1, were selected and the SDSS DR16 catalog (B. W. Lyke et al. 2020) was used as the reference. The light curves in all three bands (*g*, *r*, and *i*) are consistent with those observed by ASAS-SN and GWAC. However, the brightening phase was not detected due to the delayed start of the observations.

### 2.8. Lijiang 2.4 m Telescope

Using the Lijiang 2.4 m telescope, three high-resolution spectra of HD 251108 were obtained on 2022 November 10, November 30, and December 24, respectively. The resolution is ~32,000 at 550 nm (C. J. Wang et al. 2019). The observed spectra were reduced using the IRAF software (D. Tody 1986, 1993) following standard procedures, and then corrected to vacuum wavelength.

As shown in Figure 2, the strong H$\alpha$ emission line in all three spectra confirms the active nature of HD 251108. Notably, the spectrum obtained on November 10, two days after the X-ray flare peak, shows a significant enhancement in the Balmer lines (e.g., H$\alpha$, H$\beta$, H$\gamma$, and H$\delta$). This enhancement suggests the energy transport from the corona to the chromosphere during the flare.

## 3. Analysis and Results

### 3.1. X-Ray Spectral Fitting

Given the complexity of the flare spectra, our analysis began by exploring the possible emission components by fitting the quiescent-state spectra using a multitemperature plasma model. This initial step involved fitting the quiescent-state data to identify baseline temperature components contributing to the plasma emission under nonflaring conditions. Once the quiescent spectra were well-characterized, we proceeded to analyze the flare state by adding an additional emission component to the model.

#### 3.1.1. Quiescent State

To analyze the quiescent-state X-ray emission of HD 251108, we combined the NICER spectra collected after MJD = 59940 (observation IDs 5203530142–5203530169), during which the X-ray light curve showed relative stability. A thin thermal plasma model, `apec` (R. K. Smith et al. 2001), with one to three temperature components has been tried for the fitting of the quiescent spectrum. In these multitemperature models, the metallicity ($Z$) of different plasma components was set to be the same, and the redshift parameter was fixed to zero. As listed in Table 2, the fitting results show that the quiescent spectrum can be well represented by a 3T `apec` model (`TBabs*(apec+apec+apec)`). The unabsorbed quiescent luminosity was estimated to be $1.67^{+0.05}_{-0.04} \times 10^{32}$ erg s$^{-1}$ in 0.5–4.0 keV band, consistent with previous observations from ROSAT and eROSITA.

#### 3.1.2. Flaring State

To investigate the time-resolved spectra during the flare, we performed spectral fitting using the data from three instruments (i.e., LEIA, NICER, and Swift) with a 4T `apec` model (`TBabs*(apec+apec+apec+apec)`). Initially, the temperatures and metallicities of three components are fixed to the quiescent-state values, allowing only the emission measures (EMs) to vary freely. However, we found that the EMs of the 0.2 keV and 4.6 keV components remained close to their quiescent values, whereas the EM of the 1.2 keV component (referred to as EM$_{cool}$) continuously declined, which can be attributed to the chromospheric evaporation and condensation processes (A. O. Benz & M. Güdel 2010; H. C. Chen et al. 2022). During these processes, the plasma with a temperature around 1.2 keV is enhanced along with the higher temperature components in the corona during the flare and then eventually return to the chromosphere as the flare subsides. To avoid overfitting, we subsequently fixed the EMs of the first two components (0.2 keV and 4.6 keV) to their quiescent-state values. During the initial fittings, we also noted that the hydrogen column density ($N_H$) and the metallicity of the flare component was consistent across observations, allowing us to fix them to those of the quiescent state. The constancy of $N_H$ suggests no mass ejections along the line of sight during the flare. The best-fit parameters for each observation are listed in Tables A1–A3. In Figure 3, the flare component's EM (EM$_{hot}$) shows a notable rise and fall in sync with the light curve, reaching a peak two orders of magnitude higher than that in quiescence. In addition, the corresponding temperature ($kT_{hot}$) varied by a factor of around 3. We attempted to perform joint fitting of the spectra using quasi-simultaneous observations from three instruments during the main flare (MF) decay. This process resulted in the six data points shown in the bottom three panels of Figure 3. We note that the parameters derived from the joint fitting remain in line with those obtained from NICER observations. The NICER spectra were deemed more influential in the fitting due to their substantially higher count rates.

---

[20] https://iraf-community.github.io/pyraf.html. PyRAF is a command language for IRAF based on the Python scripting language that can be used in place of the existing IRAF CL.

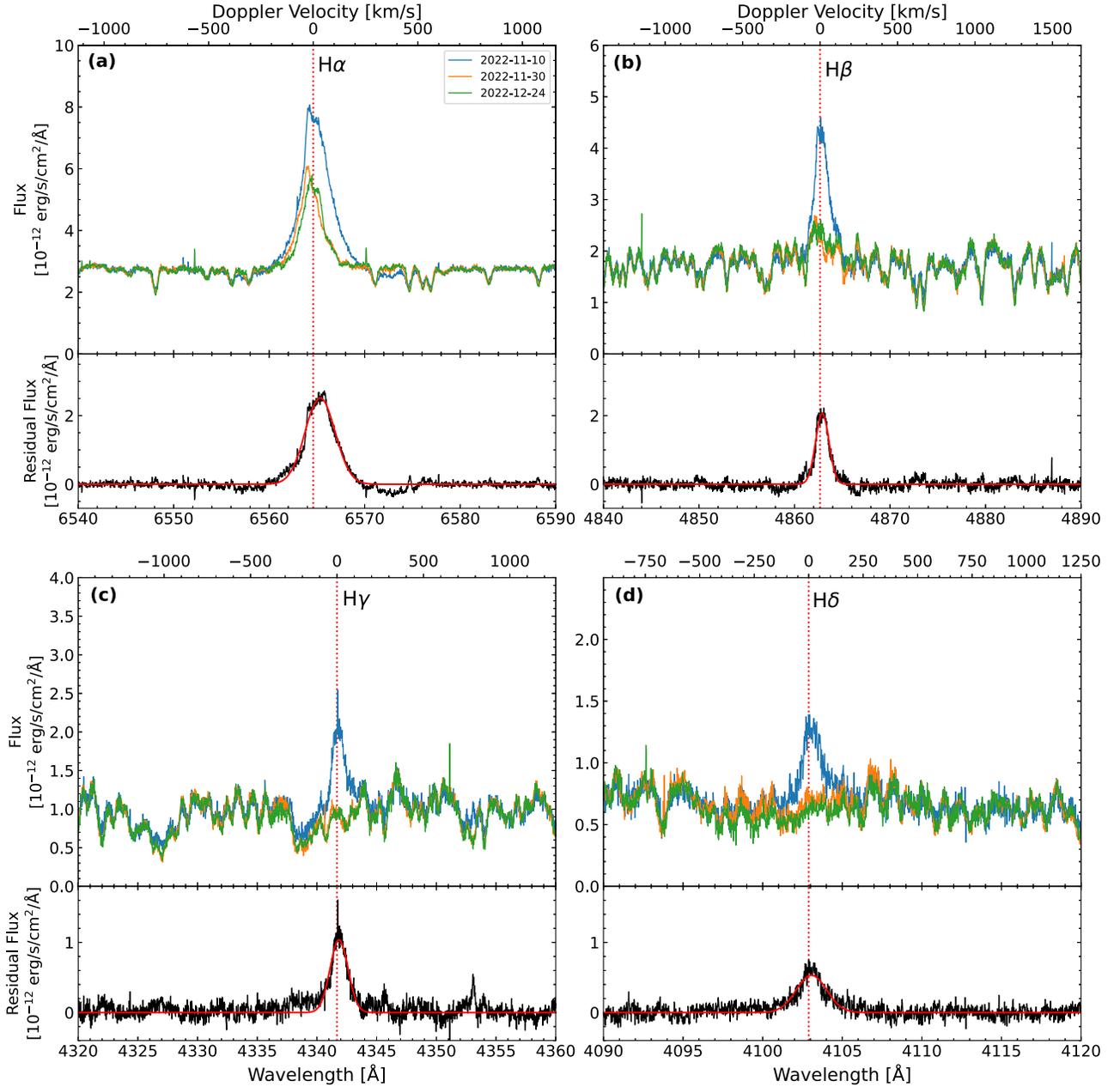

**Figure 2.** The Balmer-regions of optical spectra of HD 251108 taken by the Lijiang 2.4 m telescope. In plots (a)–(d), the residual spectra are presented in the bottom panels, each fitted by a Gaussian function (red solid line).

### 3.2. X-Ray Light Curve

Following common practice (J. C. Pandey & K. P. Singh 2012; R. Sasaki et al. 2021), the X-ray light curve was generated using photon flux corrected for absorption in 0.5–4.0 keV. Initially, the light curve is fitted using the widely adopted Fast (linear) Rise and Exponential Decay (FRED) model (R. A. Osten et al. 2016; M. N. Günther et al. 2020; R. Sasaki et al. 2021). The model is expressed as

$$c(t) = \begin{cases} c_q, & (t < t_{ST}); \\ (c_p - c_q) \times \frac{t - t_{ST}}{t_p - t_{ST}} + c_q, & (t_{ST} \leqslant t < t_p); \\ (c_p - c_q) \times \exp(-\frac{t - t_p}{\tau_d}) + c_q, & (t_p \leqslant t). \end{cases}$$

Here, $t$, $c(t)$, $t_{ST}$, $t_p$, $c_p$, and $c_q$ are time, photon flux, the time when the photon flux starts to increase, the time when the photon flux reaches the peak, the corresponding peak photon flux, and the photon flux when the star is in X-ray quiescence, respectively. $\tau_d$ is the e-folding time during the decay phase of the flares and $\tau_r = t_p - t_{ST}$ is the rise time. The time of the first LEIA observation is set to be 0. The $c_q$ was fixed to the mean value of photon flux after flare (i.e., $2.6 \times 10^{-3}$ counts cm$^{-2}$ for MJD > 59940), and the remaining four parameters ($t_{ST}$, $t_p$, $c_p$, and $\tau_d$) were derived from model fitting.

The initial fitting results show that a single FRED model cannot adequately describe the light curve during the decay phase of the main flare. The decay deviates from a pure exponential shape, with a faster decay in the early phase (MF1)

**Table 2**
Best-fit Spectral Parameters for Quiescent-state Observations

| Parameters | 1T apec | 2T apec | 3T apec |
|---|---|---|---|
| $N_H^a$ ($10^{20}$ cm$^{-2}$) | 6.43 | $5.9^{+0.1}_{-0.3}$ | $6.7^{+1.0}_{-1.6}$ |
| $kT_{q,1}^b$ (keV) | 1.57 | $1.20^{+0.03}_{-0.03}$ | $0.25^{+0.04}_{-0.03}$ |
| EM$_{q,1}^c$ ($10^{54}$ cm$^{-3}$) | 25.0 | $17.9^{+1.4}_{-1.7}$ | $4.4^{+2.3}_{-1.3}$ |
| $kT_{q,2}^b$ (keV) | ... | $8.1^{+3.6}_{-2.1}$ | $1.20^{+0.03}_{-0.04}$ |
| EM$_{q,2}^c$ ($10^{54}$ cm$^{-3}$) | ... | $6.9^{+1.0}_{-0.7}$ | $12.3^{+4.3}_{-2.4}$ |
| $kT_{q,3}^b$ (keV) | ... | ... | $4.6^{+4.7}_{-0.8}$ |
| EM$_{q,3}^c$ ($10^{54}$ cm$^{-3}$) | ... | ... | $9.4^{+1.5}_{-2.4}$ |
| $Z^d$ ($Z_\odot$) | 0.04 | $0.04^{+0.01}_{-0.01}$ | $0.07^{+0.02}_{-0.02}$ |
| $\chi^2_\nu$ (d.o.f.)$^e$ | 3.18 (128) | 1.36 (126) | 1.06 (124) |

**Notes.** All errors represent the 90% uncertainties.
[a] H$_I$ column density.
[b] Plasma temperature.
[c] EM of the plasma, calculated as EM $= 4\pi d^2 \times 10^{14}$ norm, where norm is one of the parameters in apec model.
[d] Metal abundance.
[e] Reduced $\chi^2$ ($\chi^2_\nu$) and degrees of freedom (d.o.f.)

and a slower decay in the late phase (MF2, as shown by the gray dotted line in the top panel of Figure 3). To address this, an additional FRED model was introduced for MF2, and significantly improved the fit (depicted by the red solid line in Figure 3). The light curve of secondary flare (SF) can be well fitted with a single FRED model. The best-fit parameters are showed in Table 3.

### 3.3. Flare Parameters

The flare durations are defined as $\tau_{MF} = \tau_{r,MF1} + \tau_{d,MF1} + \tau_{d,MF2} = 1090$ ks (~12.6 days) for MF and $\tau_{SF} = \tau_{r,SF} + \tau_{d,SF} = 382$ ks (~4.4 days) for SF. The total energy released by the MF in the 0.5–4.0 keV band is estimated to $E_{X,MF} = E_{X,MF1} + E_{X,MF2} = 3 \times 10^{39}$ erg, which is $10^7$ times the energy of the largest solar flares. Figure 4 shows the flare duration versus peak luminosity and flare energy, comparing these quantities with several RS CVn-type flares from previous studies (T. Tsuru et al. 1989; M. Endl et al. 1997; E. Franciosini et al. 2001; J. C. Pandey & K. P. Singh 2012; Y. Tsuboi et al. 2016; R. Sasaki et al. 2021), with X-ray luminosities and energies converted uniformly to the 0.5–4.0 keV range using webpimms with a multitemperature apec model. Additionally, we included superflares from pre-main-sequence (PMS) stars studied by K. V. Getman & E. D. Feigelson (2021), along with flares of cool stars derived from XMM-Newton data (J. P. Pye et al. 2015), although their luminosities and energies are plotted in the 0.5–8.0 keV and 0.2–12 keV bands, respectively. It is clear from Figure 4 that the MF of HD 251108 is located in the upper right-hand corner of both panels, making it the brightest, longest-lasting, and most energetic flare among all those compared, potentially representing the largest X-ray stellar flare ever detected. Notably, the parameters of the SF also rank among the highest in this context.

In addition, the stellar flares reported above as well as solar flares and microflares reported by U. Feldman et al. (1995) and T. Shimizu (1995) are plotted in the log$T$–log EM diagram (Figure 5), partitioned by isolines of flaring loop length and magnetic field strength. These isolines are derived using Equations (5) and (6) in K. Shibata & T. Yokoyama (1999).

The maximum temperature $T_{hot} \approx 86$ MK (i.e., $kT_{hot} = 7.4$ keV), derived from NICER data, serves as the lower limit for the plasma peak temperature of the main flare due to lack of constraints from LEIA observations. The diagram shows that the MF of HD 251108 has magnetic field strength and plasma temperature comparable to the RS CVn-type stellar flares while displaying one of the longest flaring loop lengths and the largest EM.

In previous studies, no additional nonthermal components have been detected in the high-energy band from stellar flares (F. Favata & J. H. M. M. Schmitt 1999; R. A. Osten et al. 2016). Thus, we can estimate the energy released across the entire X-ray range (0.1–200 keV) using the best-fitting 4T apec model in Section 3.1. Although the spectra evolve, the conversion factor from flux in the 0.5–4 keV range to the 0.1–200 keV range, derived from NICER spectra during the early decay phase of the main flare, is approximately 1.9. Therefore, the total energy released during the MF across the X-ray range is approximately $\sim 5.7 \times 10^{39}$ erg.

### 3.4. Optical Spectra

Using the three optical spectra obtained by Lijiang 2.4 m telescope, we analyzed the emission line properties during the flares. To obtain the residual spectrum, we subtracted the average of the later two quiescent spectra from the first (flare) spectrum. The residual profiles of four Balmer lines were fitted with single-Gaussian model (see bottom panels of Figure 2, plots (a)–(d)). The derived parameters are presented in Table A4. The maximum projected velocity $V_{max}$, was calculated based on the minimum and maximum wavelengths where the residual profile exceeds $1\sigma$ above the continuum. The high $V_{max}$ values suggest fast-moving chromospheric plasma, possibly driven by energy injections from flare-accelerated electrons.

### 4. Discussion

#### 4.1. Cooling Process and Flare Loop Geometry

As mentioned in Section 3.2, the X-ray light curve of the MF shows a double-exponential decay. Such a decay is commonly seen in many stellar flares (F. Reale et al. 2004; K. V. Getman et al. 2008; B. J. Wargelin et al. 2008), which is believed to be similar to some large solar flares (J. Lin & T. G. Forbes 2000; M. J. Aschwanden & D. Alexander 2001; J. Lin 2002; L. K. Kashapova et al. 2021). The flare process is typically explained as follows (F. Reale et al. 2004). Initially, magnetic reconnection in an active region on the stellar surface triggers a heat pulse, igniting a large magnetic loop and causing the initial flare burst. Once the first heat pulse subsides, the flare quickly fades, but residual heat ignites a secondary system of loops, akin to an arcade of loops observed in solar flares. The cooling of these loops leads to the slower decay observed later. The constant $T_{hot}$ before and around the time of the secondary peak ($t_{p,MF2}$) supports the idea that the event is due to the cooling of an established loop system rather than a brand-new loop causing a subsequent, weaker flare. A similar conclusion was discussed in F. Reale et al. (2004). In this scenario, the flare duration, temperature, and EM obtained in Section 3 can be used to infer the geometry of the magnetic loops and the physical properties of the flaring plasma.

The detailed calculations and results for the loop parameters are presented in Appendix B. The magnetic field strength and

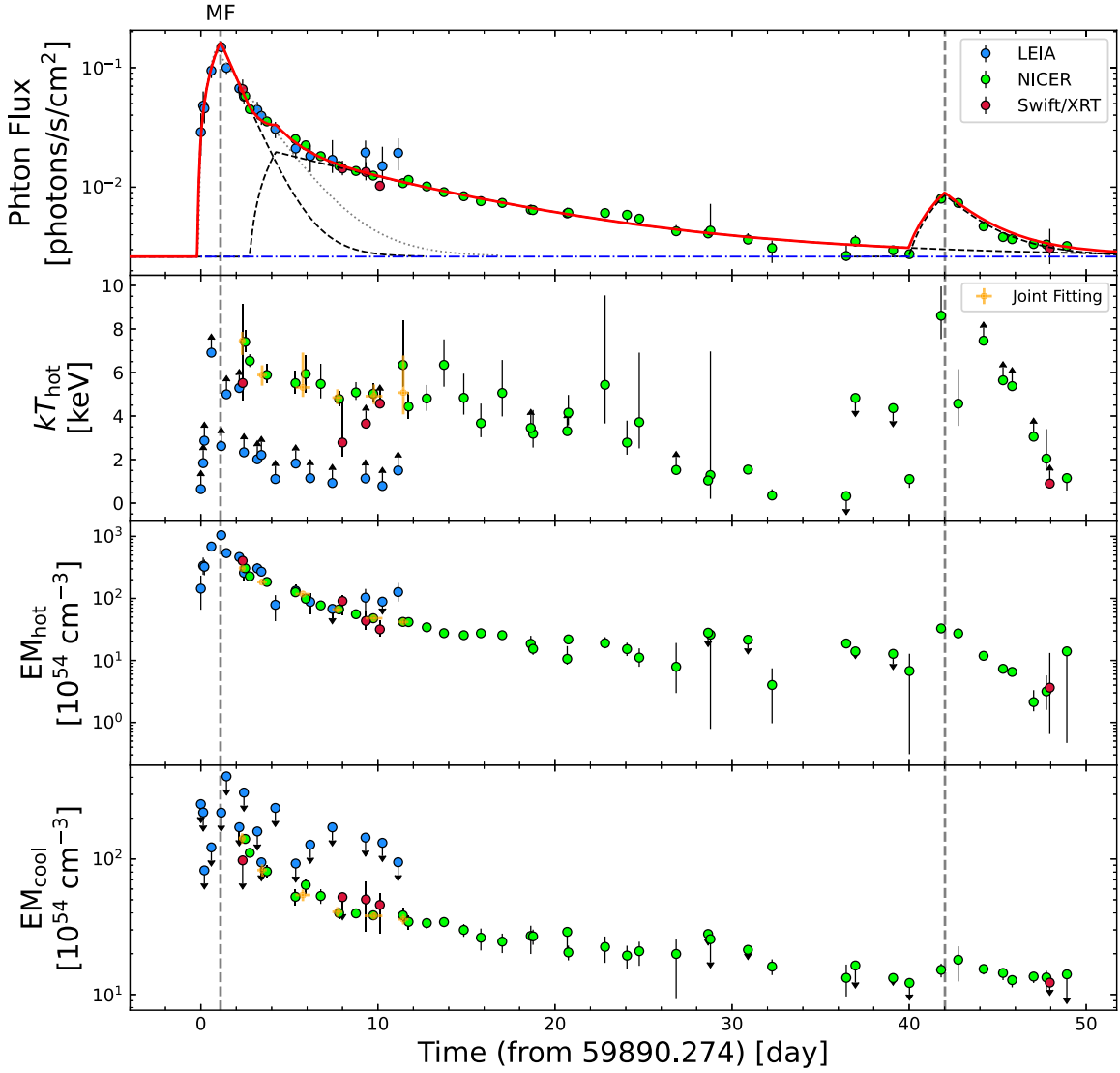

**Figure 3.** X-ray light curve and evolution of spectral parameters from LEIA, NICER, and Swift/XRT spectra. From top to bottom, it shows the X-ray photon flux light curve in the 0.5–4.0 keV, the time variations in the temperature and EM of the flare plasma component, and the time variation in the EM of the 1.2 keV plasma component. The gray dashed lines are consistent with those in Figure 1. In the top panel, each black dashed line represents a FRED model employed to fit the light curve. The blue dotted–dashed line indicates the quiescent photon flux, while the red solid line shows the combined fitting result. The gray dotted line illustrates the single FRED fitting of the MF, which does not adequately describe the decay phase. The joint fitting parameters of nearly simultaneous observations from different instruments are also displayed in the bottom three panels.

Table 3
Best-fit Parameters of the X-Ray Light Curve

| | $t_p^a$ (Days Since MJD = 59890.274) | $c_p^b$ (Counts s$^{-1}$ cm$^{-2}$) | $\tau_r^c$ (ks) | $\tau_d^d$ (ks) | $L_{X,p}^e$ (10$^{32}$ erg s$^{-1}$) | $f_X^f$ (10$^{-5}$ erg cm$^{-2}$) | $E_X^g$ (10$^{38}$ erg) |
|---|---|---|---|---|---|---|---|
| MF1 | 1.1 ± 0.3 | 0.16 ± 0.04 | 112 ± 50 | 109 ± 9 | 110.8 ± 29.7 | 6.0 ± 1.9 | 18.3 ± 5.7 |
| MF2 | 4.2 ± 0.2 | 0.020 ± 0.001 | 128 ± 12 | 869 ± 34 | 13.3 ± 1.7 | 4.1 ± 1.1 | 12.4 ± 1.7 |
| SF  | 42.0 ± 0.2 | 0.009 ± 0.001 | 177 ± 18 | 205 ± 24 | 5.8 ± 0.8 | 0.6 ± 0.1 | 1.7 ± 0.3 |

**Notes.** All errors represent the 90% uncertainties.
[a] Flare peak time since the observation start time of the first detection by LEIA.
[b] Peak photon flux in the 0.5–4.0 keV band.
[c] Flare rise time, which is the difference between the flare start time and its peak time (namely, $t_p - t_{ST}$).
[d] Flare e-folding decay time.
[e] Flare peak luminosity in 0.5–4.0 keV with absorption corrected, calculated by $L_{X,p} = 4\pi d^2 F_{X,p}$, where $d$ is the distance of HD 251108 and $F_{X,p}$ is the flare peak flux, assumed to be isotropic.
[f] Time-integrated X-ray fluence in 0.5–4.0 keV of the flare calculated by $f_X = F_{X,p}(\tau_r/2 + \tau_d)$.
[g] Total energy released during the flare, i.e., $E_X = L_{X,p}(\tau_r/2 + \tau_d) = 4\pi d^2 f_X$.

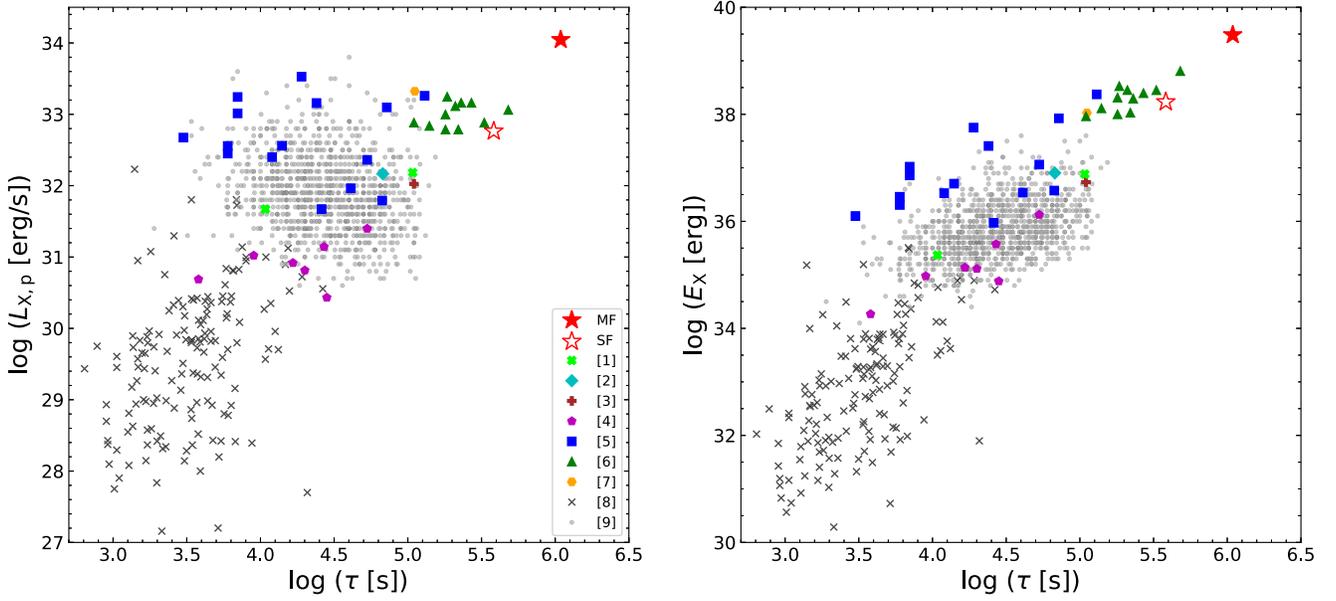

**Figure 4.** Flare duration ($\tau$) vs. flare peak luminosity ($L_{X,p}$, left-hand panel) and flare energy ($E_X$, right-hand panel) for MF and SF, along with stellar flares reported by [1]: T. Tsuru et al. (1989), [2]: M. Endl et al. (1997), [3]: E. Franciosini et al. (2001), [4]: J. C. Pandey & K. P. Singh (2012), [5]: Y. Tsuboi et al. (2016), [6]: R. Sasaki et al. (2021), [7]: S. Karmakar et al. (2023), [8]: J. P. Pye et al. (2015), and [9]: K. V. Getman & E. D. Feigelson (2021). The flares from [1] to [7] all took place in RS CVn-type binary systems, with the energy band converted into 0.5–4.0 keV. The energy bands are 0.2–12 keV in [8] and 0.5–8.0 keV in [9], respectively.

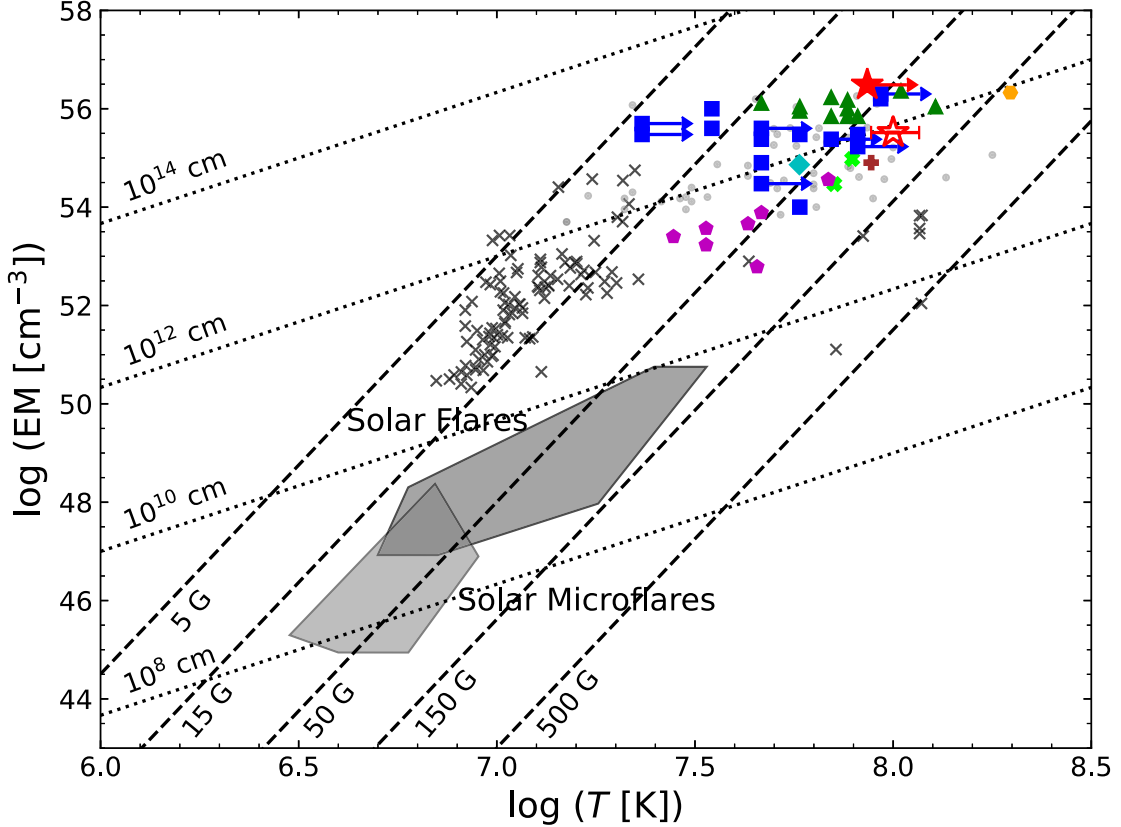

**Figure 5.** EM vs. plasma temperature ($T$). The symbols are the same as those in Figure 4, with the exception that the gray points are from K. V. Getman et al. (2021), which represent 55 bright superflares selected from K. V. Getman & E. D. Feigelson (2021). Plus filled polygons symbolized solar flares (U. Feldman et al. 1995) and solar microflares (T. Shimizu 1995). The dashed lines indicate the EM–T relation (EM $\propto B^{-5}T^{17/2}$) for a constant magnetic field, and the dotted lines show the EM–T relation (EM $\propto L^{5/3}T^{8/3}$) constrained by certain loop lengths (K. Shibata & T. Yokoyama 1999).

flaring loop length derived from the cooling timescale of MF align with those indicated by the the EM–T diagram in Figure 5. The estimated loop length is approximately $1.9 \times 10^{12}$ cm.

A comparison with eight flares from RS CVn-type stars reported by J. C. Pandey & K. P. Singh (2012) shows that the estimated loop length $L$ in our work is at least an order of magnitude larger. According to the RTV scaling law, where $p \propto L^{-1}T^3$ (see Appendix B), this results in a much smaller pressure and, consequently, a weaker magnetic field. The longevity of such an extensive loop structure may primarily be due to the star's low gravitational acceleration. The estimated magnetic field strength and half-length of flaring loop are also consistent with Figure 5.

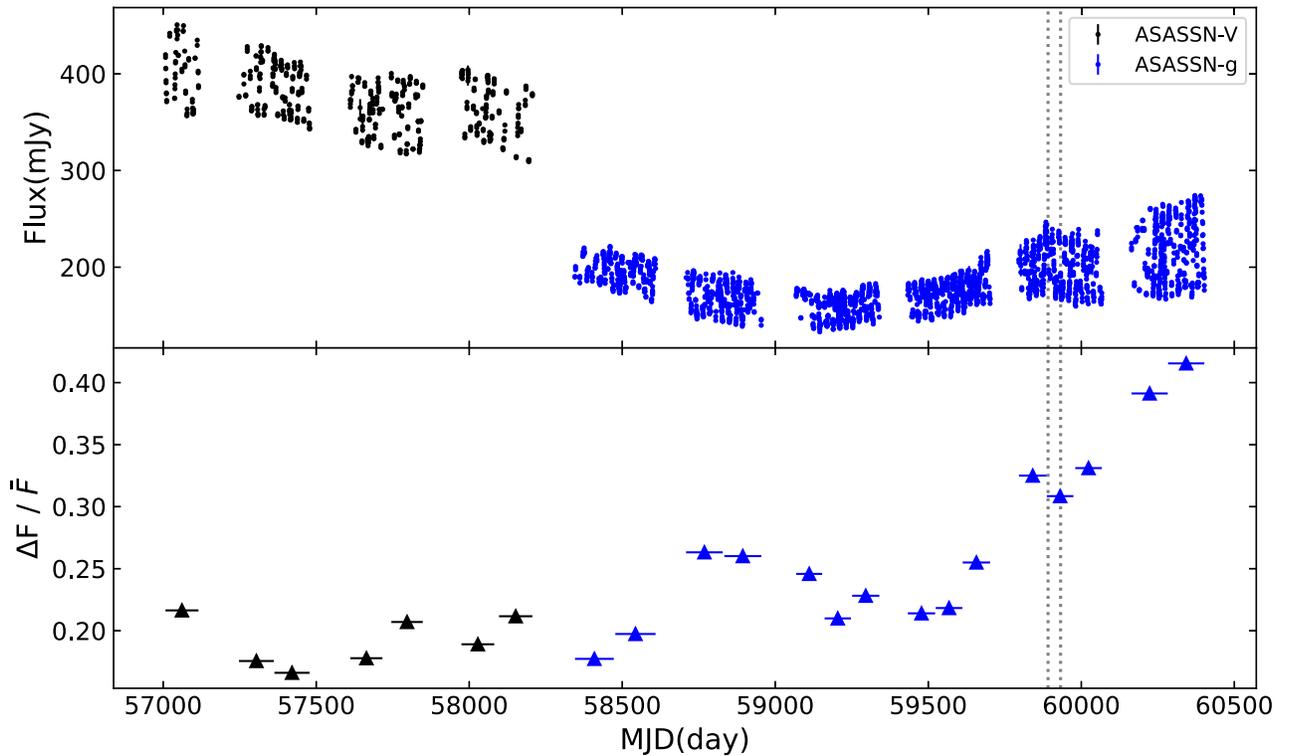

**Figure 6.** The entire light curves from ASAS-SN *V*-band and *g*-band (top panel) and the temporal evolution of $\Delta F/\bar{F}$ (bottom panel). The gray dotted lines mark the time MF and SF occur.

It is important to note that the loop size may be greatly underestimated because the actual peak temperature $kT_{\rm hot,peak}$ could significantly exceed 7.4 keV. For instance, assuming $kT_{\rm hot,peak} = 10$ keV yields a loop length of approximately $2.2 \times 10^{12}$ cm, while $kT_{\rm hot,peak} = 20$ keV results in a length of $\sim 3.2 \times 10^{12}$ cm (see Table B1). Additionally, the choice of energy band can influence the shape of the light curve, affecting the derived cooling timescales and loop parameters. We therefore attempted to fit the light curve in the 0.1–200 keV range (based on the best-fitting spectral models), again assuming $kT_{\rm hot,peak} = 10\text{–}20$ keV. The results suggest that the loop length could increase by 10%–30%. Note that several other factors, such as sustained heating (F. Reale et al. 1997), may also influence the estimation of the loop length. A more comprehensive analysis of the flaring processes is required, but such an analysis is beyond the scope of this work and will be addressed in future research.

### 4.2. Stellar Activity

We used the photometric data from ASAS-SN Sky Patrol v1.0 to examine the magnetic activity of HD 251108. The light curve of HD 251108 shows a rotational modulation with a period of $\approx 21.21$ days (M. Kiraga 2012), which is also visible in Figure 1. We divided the ASAS-SN light curve, spanning from 2014 to 2024, into 22 epochs (seven epochs for the *V*-band and 15 epochs for the *g*-band), and calculated a flux variation ratio $\Delta F/\bar{F}$ for each epoch, defined as follows,

$$\Delta F/\bar{F} = \frac{F_{95\%} - F_{5\%}}{\bar{F}},$$

where $\bar{F}$ is the mean value of flux and $F_{95\%}$ and $F_{5\%}$ are the 95th and 5th percentile flux, respectively. The long-term light curve shows significant variations in both flux and rotation-modulated amplitude, which may reveal long-term evolution of large star spots in active regions on stellar surface. (Figure 6). The flare occurred when the amplitude of the modulation increased to a high value, suggesting an expansion of starspot area and hence a higher stellar activity.

We examined the magnetic activity of HD 251108 by introducing the X-ray luminosity to bolometric luminosity ratio, $R_{\rm X} = L_{\rm X}/L_{\rm bol}$ (R. Pallavicini et al. 1981). Using the bolometric luminosity estimated from stellar parameters mentioned in Section 1 and the quiescent X-ray luminosity derived from NICER, we derived $R_{\rm X} \sim 5 \times 10^{-4}$, placing HD 251108 within the saturated region ($R_{\rm X} \sim 10^{-3}$), indicative of high X-ray activity (N. J. Wright et al. 2011).

It is noteworthy that although the MF lasts for about 40 days, much longer than the rotational period of the star, the light curve seems to decay smoothly and be unaffected by stellar rotation. Here, we proposed two possible explanations. First, using stellar parameters provided in Section 1, we determined stellar radius to be $15.2\,R_\odot$, indicating that the flaring loop is $\sim 1.8$ times stellar radius. Therefore, the main body of the loops can remain visible even when the footpoints are shielded by the star. Alternatively, the flares may occur in the polar magnetic active region which can be always seen in the line of sight.

## 5. Summary

We reported a superflare event, which happened on a nearby RS CVn candidate HD 251108, was detected by LEIA, and followed by multiwavelength observations. The detected X-ray flux rise lasted for nearly 40 days with peak X-ray luminosity estimated to be $\sim 1.1 \times 10^{34}$ erg s$^{-1}$ in 0.5–4.0 keV, making it the longest-lasting and potentially the most luminous stellar X-ray flare. A change of e-folding timescale was found during the flare decay phase, thus a double-FRED model was chosen to fit the X-ray light curve. The fit result indicated that the main energy release occurred in the fast decay phase. The time-resolved X-ray spectra were well fitted by the four-temperature `apec` model with three components frozen as quiescent background. The derived flare parameters ($T_{hot}$, $EM_{hot}$, duration and peak luminosity) are all located at the upper end of the parameter spaces constructed by superflares reported in previous studies. The flaring loop length is roughly estimated to be $\sim 1.9 \times 10^{12}$ cm, and the magnetic field strength $B$ is about 55 G. The derived energy release in 0.5–4.0 keV is $\sim 3 \times 10^{39}$ erg, which suggests this is possibly the most energetic ever detected X-ray stellar flare.

Such an energetic flare is actually not unexpected since the magnetic activity of HD 251108 is rather high ($R_X \approx 5 \times 10^{-4}$). In addition, a long-term variability is evident in the ASAS-SN photometric data. The flare occurred during an epoch when the amplitude of the light curve variation significantly increased, indicating an expansion of the starspot area and hence enhanced stellar activity. Apart from the intense magnetic activities that foster conditions for the superflare, the low surface gravity of HD 251108 also contributes to the prolonged duration of this flare.

With the launch of EP on 2024 January 9, which monitors the X-ray sky with much larger FoV and wider range of timescales than LEIA, it is expected to detect a massive amount of stellar flares, which can fundamentally deepen our understanding of this explosive phenomenon.

## Acknowledgments

We thank the anonymous referee for their very helpful comments and suggestions that improved this paper. This work is based on the data obtained with LEIA, a pathfinder of the Einstein Probe mission, which is supported by the Strategic Priority Program on Space Science of Chinese Academy of Sciences (grant Nos. XDA15310000, XDA15052100). We acknowledge the support by the National Key Research and Development Programs of China (NKRDPC) under grant Nos. 2022YFF0711404, 2021YFA0718500, 2022SKA0130102, 2022SKA0130100, 2019YFA0405504, and 2019YFA0405000, National Natural Science Foundation of China (NSFC) under grant Nos. 12103061, 11988101, 12273057, 12203071, 11833002, 12090042, 12103047, and 12473016, and the Strategic Priority Program on Space Science of Chinese Academy of Sciences (grant Nos. XDB41000000, XDB0550200). A.J.C.T. acknowledges support from the Spanish Ministry projects PID2020-118491GB-I00 and PID2023-151905OB-I00 and Junta de Andalucìa grant P20_010168 and from the Severo Ochoa grant CEX2021-001131-S funded by MCIN/AEI/ 10.13039/ 501100011033. This work also made use of data supplied by NASA through the NICER team and Swift team, and software provided by the High Energy Astrophysics Science Archive Research Center (HEASARC), which is a service of the Astrophysics Science Division at NASA/GSFC. We acknowledge use of the VizieR catalog access tool, operated at CDS, Strasbourg, France, and of Astropy, a community-developed core Python package for Astronomy (Astropy Collaboration 2013). This research made use of Photutils, an Astropy package for detection and photometry of astronomical sources.

*Facilities:* EP (W. Yuan et al. 2022), NICER, Swift, SVOM, BOOTES, YAO:2.4m.

*Software*: Python, Astropy, Heasoft (K. A. Arnaud 1996), IRAF.

## Appendix A
## Best-fit Spectral Parameters

All best-fit parameters of X-ray and optical spectra discussed in Sections 3.1–3.4 are listed in this appendix. In addition, spectral fitting for LEIA and NICER are illustrated in Figure A1.

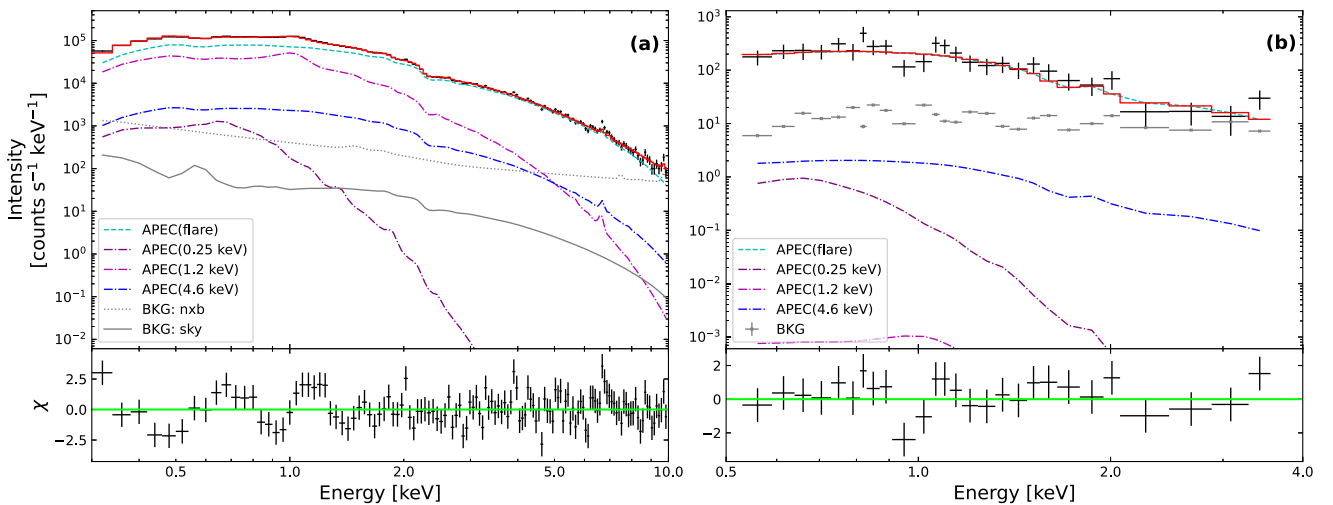

**Figure A1.** Two examples of X-ray spectra during MF, fitted by 4T `apec` model (red solid line) composed of a flare component (cyan dashed) and three quiescent components (dashdot). (a) The NICER spectra is corresponding to the No.01 observation in Table A1, and (b) the LEIA spectra is corresponding to the No.05 observation in Table A2.

Table A1
Best-fit Spectral Parameters for Each NICER Observation

| No. | Obs. Start Time (UTC) | Exposure Time (s) | $kT_{\rm hot}^{a}$ (keV) | $EM_{\rm hot}^{b}$ ($10^{54}$ cm$^{-3}$) | $EM_{\rm cool}^{c}$ ($10^{54}$ cm$^{-3}$) | $L_{\rm X}$ ($10^{32}$ erg s$^{-1}$) | $\chi_\nu^2$ (d.o.f.) | Joint No.[d] |
|---|---|---|---|---|---|---|---|---|
| 01 | 2022-11-09 18:37:00 | 2185 | $7.4^{+0.5}_{-0.5}$ | $306.4^{+9.4}_{-8.9}$ | $140.5^{+9.8}_{-10.2}$ | $38.7^{+0.8}_{-0.7}$ | 1.47 (129) | 1 |
| 02 | 2022-11-10 00:49:40 | 7199 | $6.5^{+0.3}_{-0.3}$ | $227.3^{+6.3}_{-6.1}$ | $111.0^{+6.2}_{-6.3}$ | $29.2^{+0.6}_{-0.5}$ | 1.41 (136) | ⋯ |
| 03 | 2022-11-11 00:04:00 | 2035 | $5.9^{+0.5}_{-0.4}$ | $184.2^{+7.2}_{-7.2}$ | $80.9^{+8.2}_{-7.8}$ | $23.3^{+0.5}_{-0.5}$ | 1.36 (120) | 2 |
| 04 | 2022-11-12 14:47:00 | 1771 | $5.5^{+0.6}_{-0.5}$ | $125.8^{+6.3}_{-6.0}$ | $52.6^{+6.9}_{-7.0}$ | $15.9^{+0.3}_{-0.3}$ | 1.18 (114) | 3 |
| 05 | 2022-11-13 04:44:00 | 1101 | $5.9^{+0.9}_{-0.9}$ | $98.8^{+7.5}_{-5.8}$ | $64.3^{+7.0}_{-8.8}$ | $14.1^{+0.3}_{-0.3}$ | 0.92 (113) | 3 |
| 06 | 2022-11-14 01:02:00 | 1221 | $5.5^{+0.9}_{-0.7}$ | $77.1^{+5.8}_{-5.5}$ | $53.3^{+6.6}_{-6.6}$ | $11.4^{+0.3}_{-0.2}$ | 1.28 (109) | ⋯ |
| 07 | 2022-11-15 02:00:01 | 5719 | $4.8^{+0.4}_{-0.3}$ | $66.4^{+3.2}_{-3.0}$ | $39.9^{+3.3}_{-3.5}$ | $9.5^{+0.2}_{-0.2}$ | 1.44 (123) | 4 |
| 08 | 2022-11-16 00:45:41 | 9415 | $5.1^{+0.5}_{-0.3}$ | $55.6^{+2.5}_{-2.7}$ | $39.7^{+3.1}_{-2.8}$ | $8.5^{+0.2}_{-0.2}$ | 1.27 (130) | ⋯ |
| 09 | 2022-11-17 00:11:00 | 9914 | $5.0^{+0.4}_{-0.4}$ | $47.8^{+2.3}_{-2.3}$ | $38.4^{+2.7}_{-2.6}$ | $7.6^{+0.2}_{-0.1}$ | 1.39 (131) | 5 |
| 10 | 2022-11-18 16:22:20 | 2018 | $6.3^{+2.0}_{-1.0}$ | $41.8^{+3.4}_{-3.6}$ | $38.4^{+4.8}_{-4.1}$ | $7.3^{+0.2}_{-0.2}$ | 1.15 (112) | 6 |
| 11 | 2022-11-19 00:02:20 | 3553 | $4.4^{+0.7}_{-0.5}$ | $41.5^{+3.8}_{-3.4}$ | $34.4^{+3.9}_{-4.2}$ | $6.8^{+0.1}_{-0.1}$ | 1.08 (114) | 6 |
| 12 | 2022-11-20 00:48:00 | 6257 | $4.8^{+0.6}_{-0.6}$ | $34.2^{+2.8}_{-2.3}$ | $33.7^{+2.7}_{-3.1}$ | $6.1^{+0.1}_{-0.1}$ | 1.13 (117) | ⋯ |
| 13 | 2022-11-21 00:05:20 | 4979 | $6.3^{+1.2}_{-1.0}$ | $27.5^{+2.2}_{-1.8}$ | $34.3^{+2.4}_{-2.8}$ | $5.6^{+0.1}_{-0.1}$ | 1.07 (116) | ⋯ |
| 14 | 2022-11-22 02:47:40 | 3273 | $4.8^{+1.1}_{-0.8}$ | $25.5^{+2.9}_{-2.6}$ | $30.0^{+3.1}_{-3.3}$ | $5.1^{+0.1}_{-0.1}$ | 1.22 (112) | ⋯ |
| 15 | 2022-11-23 02:01:40 | 1394 | $3.7^{+0.9}_{-0.6}$ | $27.4^{+4.9}_{-4.1}$ | $26.2^{+4.4}_{-5.1}$ | $4.9^{+0.1}_{-0.1}$ | 1.21 (102) | ⋯ |
| 16 | 2022-11-24 07:00:40 | 2107 | $5.1^{+1.5}_{-1.1}$ | $25.5^{+3.9}_{-3.0}$ | $24.6^{+3.6}_{-4.3}$ | $4.8^{+0.1}_{-0.1}$ | 0.81 (108) | ⋯ |
| 17 | 2022-11-25 21:41:40 | 563 | $>3.5$ | $18.5^{+6.4}_{-3.8}$ | $27.1^{+5.1}_{-7.2}$ | $4.3^{+0.1}_{-0.1}$ | 1.30 (94) | ⋯ |
| 18 | 2022-11-26 00:47:40 | 3696 | $3.2^{+1.1}_{-0.6}$ | $15.4^{+3.6}_{-3.1}$ | $26.9^{+3.3}_{-3.6}$ | $3.8^{+0.1}_{-0.1}$ | 1.24 (118) | ⋯ |
| 19 | 2022-11-27 23:07:19 | 773 | $>3.3$ | $10.6^{+6.4}_{-2.0}$ | $29.0^{+2.5}_{-7.7}$ | $3.6^{+0.1}_{-0.1}$ | 0.96 (91) | ⋯ |
| 20 | 2022-11-28 00:40:12 | 4718 | $4.2^{+0.8}_{-0.6}$ | $21.8^{+2.4}_{-2.2}$ | $20.4^{+2.5}_{-2.6}$ | $4.1^{+0.1}_{-0.1}$ | 1.16 (117) | ⋯ |
| 21 | 2022-11-30 02:23:00 | 1074 | $5.4^{+4.1}_{-1.8}$ | $18.9^{+4.8}_{-3.3}$ | $22.5^{+4.3}_{-5.3}$ | $4.1^{+0.1}_{-0.1}$ | 0.89 (101) | ⋯ |
| 22 | 2022-12-01 08:06:20 | 3508 | $2.8^{+1.0}_{-0.6}$ | $15.2^{+4.1}_{-3.5}$ | $19.4^{+3.5}_{-4.0}$ | $3.2^{+0.1}_{-0.1}$ | 1.05 (118) | ⋯ |
| 23 | 2022-12-02 00:47:00 | 1383 | $3.7^{+3.2}_{-1.2}$ | $11.1^{+4.5}_{-3.2}$ | $20.9^{+3.6}_{-4.6}$ | $3.2^{+0.1}_{-0.1}$ | 1.34 (104) | ⋯ |
| 24 | 2022-12-04 02:44:40 | 547 | $>1.5$ | $7.9^{+11.2}_{-4.9}$ | $19.9^{+5.6}_{-10.7}$ | $2.8^{+0.1}_{-0.1}$ | 1.19 (91) | ⋯ |
| 25 | 2022-12-05 22:00:40 | 1066 | $1.0$[e] | $<28.0$ | $<28.0$ | $2.5^{+0.2}_{-0.2}$ | 1.00 (80) | ⋯ |
| 26 | 2022-12-06 01:00:20 | 492 | $1.3^{+5.7}_{-1.1}$ | $26.0^{+1.9}_{-25.2}$ | $<25.6$ | $2.6^{+0.1}_{-0.1}$ | 1.32 (89) | ⋯ |
| 27 | 2022-12-08 02:42:40 | 1033 | $1.5$[e] | $<21.5$ | $<21.4$ | $2.1^{+0.1}_{-0.1}$ | 1.23 (101) | ⋯ |
| 28 | 2022-12-09 12:44:00 | 460 | $0.4^{+0.3}_{-0.2}$ | $4.0^{+3.4}_{-3.1}$ | $16.1^{+2.1}_{-2.2}$ | $2.0^{+0.1}_{-0.1}$ | 0.93 (92) | ⋯ |
| 29 | 2022-12-13 17:19:21 | 625 | $<0.3$ | 18.8 | $13.3^{+3.4}_{-3.6}$ | $1.4^{+0.2}_{-0.2}$ | 1.21 (77) | ⋯ |
| 30 | 2022-12-14 05:42:21 | 747 | $<4.8$ | $<14.0$ | $<16.4$ | $1.8^{+0.1}_{-0.1}$ | 1.44 (87) | ⋯ |
| 31 | 2022-12-16 08:49:06 | 3210 | $<4.4$ | $<12.8$ | $<13.2$ | $1.7^{+0.1}_{-0.1}$ | 1.16 (108) | ⋯ |
| 32 | 2022-12-17 06:46:00 | 3129 | $1.1^{+0.1}_{-0.4}$ | $6.8^{+6.0}_{-6.4}$ | $<12.2$ | $1.7^{+0.1}_{-0.1}$ | 1.24 (105) | ⋯ |
| 33 | 2022-12-19 01:51:27 | 7232 | $8.6^{+1.3}_{-1.1}$ | $32.8^{+1.4}_{-1.2}$ | $15.2^{+1.7}_{-1.8}$ | $5.0^{+0.1}_{-0.1}$ | 1.07 (119) | ⋯ |
| 34 | 2022-12-20 01:05:06 | 780 | $4.6^{+1.6}_{-1.0}$ | $27.3^{+5.2}_{-4.0}$ | $18.0^{+4.6}_{-5.6}$ | $4.5^{+0.1}_{-0.1}$ | 1.37 (99) | ⋯ |
| 35 | 2022-12-21 11:27:00 | 3376 | $>7.5$ | $11.9^{+1.0}_{-0.9}$ | $15.4^{+1.0}_{-1.4}$ | $3.0^{+0.1}_{-0.1}$ | 1.03 (113) | ⋯ |
| 36 | 2022-12-22 13:47:00 | 4102 | $>5.6$ | $7.3^{+1.1}_{-0.8}$ | $14.4^{+0.9}_{-1.6}$ | $2.5^{+0.1}_{-0.1}$ | 1.14 (119) | ⋯ |
| 37 | 2022-12-23 02:10:40 | 4919 | $>5.4$ | $6.5^{+1.1}_{-0.8}$ | $12.8^{+1.0}_{-1.5}$ | $2.3^{+0.1}_{-0.1}$ | 1.11 (119) | ⋯ |
| 38 | 2022-12-24 07:16:59 | 7324 | $>3.1$ | $2.1^{+1.2}_{-0.6}$ | $13.6^{+0.8}_{-1.4}$ | $1.9^{+0.1}_{-0.1}$ | 1.17 (122) | ⋯ |
| 39 | 2022-12-25 00:39:00 | 7514 | $2.0^{+1.4}_{-0.5}$ | $3.2^{+2.6}_{-1.6}$ | $13.4^{+1.6}_{-2.5}$ | $1.9^{+0.1}_{-0.1}$ | 1.50 (118) | ⋯ |
| 40 | 2022-12-26 01:11:20 | 3045 | $1.1^{+0.1}_{-0.6}$ | $14.0^{+0.6}_{-13.5}$ | $<14.1$ | $1.7^{+0.1}_{-0.1}$ | 1.54 (119) | ⋯ |

**Notes.** All errors represent the 90% uncertainties.
[a] Plasma temperature of the flare component.
[b] EM of the flare component.
[c] EM of the 1.2 keV component.
[d] The serial number indicating which joint fitting the spectrum was used for.
[e] The upper and lower limits could not be determined.

**Table A2**
Best-fit Spectral Parameters for Each LEIA Observation

| No. | Obs. Start Time (UTC) | Exposure Time (s) | $kT_{hot}$ keV | $EM_{hot}$ $10^{54}$ cm$^{-3}$ | $EM_{cool}$ $10^{54}$ cm$^{-3}$ | $L_X$ $10^{32}$ erg s$^{-1}$ | $\chi_\nu^2$ (d.o.f.) | Joint No.[a] |
|---|---|---|---|---|---|---|---|---|
| 01 | 2022-11-07 06:26:36 | 1115 | >0.6 | $144^{+90}_{-78}$ | <254 | $13.5^{+8.2}_{-6.6}$ | 1.11 (37) | ... |
| 02 | 2022-11-07 09:35:57 | 1116 | >1.8 | $336^{+120}_{-96}$ | <220 | $33.5^{+11.1}_{-10.7}$ | 0.72 (37) | ... |
| 03 | 2022-11-07 11:10:37 | 1084 | >2.9 | $324^{+98}_{-88}$ | <82 | $32.2^{+9.4}_{-8.0}$ | 0.93 (44) | ... |
| 04 | 2022-11-07 19:04:00 | 2150[b] | >6.9 | $685^{+94}_{-95}$ | <122 | $68.2^{+7.7}_{-10.5}$ | 1.33 (22) | ... |
| 05 | 2022-11-08 09:16:03 | 2230[b] | >2.6 | $1045^{+158}_{-124}$ | <220 | $104.1^{+8.4}_{-16.2}$ | 0.90 (26) | ... |
| 06 | 2022-11-08 15:34:44 | 3348[b] | >5.0 | $538^{+78}_{-87}$ | <407 | $65.4^{+9.1}_{-9.7}$ | 1.00 (32) | ... |
| 07 | 2022-11-09 08:56:08 | 3344[b] | >5.3 | $465^{+59}_{-70}$ | <172 | $47.9^{+5.0}_{-6.5}$ | 1.33 (29) | 1 |
| 08 | 2022-11-09 15:14:49 | 2919[b] | >2.3 | $259^{+55}_{-63}$ | <309 | $33.9^{+5.9}_{-6.0}$ | 0.53 (23) | 1 |
| 09 | 2022-11-10 10:10:53 | 2231[b] | >2.0 | $305^{+66}_{-52}$ | <160 | $30.4^{+4.9}_{-6.7}$ | 0.86 (15) | 2 |
| 10 | 2022-11-10 14:54:53 | 3346[b] | >2.2 | $270^{+57}_{-43}$ | <95 | $28.2^{+3.0}_{-5.4}$ | 1.02 (20) | 2 |
| 11 | 2022-11-11 08:16:16 | 4149[b] | >1.1 | $79^{+34}_{-36}$ | <238 | $14.6^{+3.7}_{-2.1}$ | 0.54 (22) | ... |
| 12 | 2022-11-12 11:04:04 | 5453[b] | >1.8 | $133^{+33}_{-26}$ | <93 | $14.2^{+2.8}_{-3.6}$ | 0.74 (21) | 3 |
| 13 | 2022-11-13 09:09:23 | 3243[b] | >1.1 | $88^{+30}_{-33}$ | <127 | $12.5^{+3.4}_{-4.6}$ | 0.86 (10) | 3 |
| 14 | 2022-11-14 16:42:59 | 1078 | >0.9 | $12^{+56}_{-0}$ | <171 | $8.9^{+6.2}_{-2.8}$ | 0.95 (31) | 4 |
| 15 | 2022-11-16 12:52:49 | 2150[b] | >1.1 | $103^{+40}_{-43}$ | <144 | $12.5^{+4.9}_{-4.8}$ | 0.79 (6) | 5 |
| 16 | 2022-11-17 12:32:31 | 1109 | >0.8 | <89 | <132 | $7.0^{+3.3}_{-2.5}$ | 1.23 (26) | 5 |
| 17 | 2022-11-18 09:02:56 | 2175[b] | >1.5 | $127^{+51}_{-38}$ | <95 | $14.1^{+3.2}_{-4.7}$ | 0.96 (6) | 6 |

**Notes.** All errors represent the 90% uncertainties.
[a] The serial number indicating which joint fitting the spectrum was used for.
[b] Data from consecutive observations are combined for better spectral analysis.

**Table A3**
Best-fit Spectral Parameters for Each Swift/XRT Observation

| No. | Obs. Start Time (UTC) | Exposure Time (s) | $kT_{hot}$ (keV) | $EM_{hot}$ ($10^{54}$ cm$^{-3}$) | $EM_{cool}$ ($10^{54}$ cm$^{-3}$) | $L_X$ ($10^{32}$ erg s$^{-1}$) | $\chi_\nu^2$ (d.o.f.) | Joint No.[a] |
|---|---|---|---|---|---|---|---|---|
| 01 | 2022-11-09 15:04:35 | 1499 | $5.5^{+9.1}_{-4.7}$ | $403.9^{+21.5}_{-74.2}$ | <97.7 | $39.2^{+1.8}_{-1.8}$ | 0.81 (130) | 1 |
| 02 | 2022-11-15 06:21:44 | 541 | $2.8^{+4.7}_{-2.1}$ | $90.8^{+22.4}_{-37.1}$ | <52.2 | $9.1^{+0.7}_{-0.7}$ | 1.08 (59) | 4 |
| 03 | 2022-11-16 14:12:37 | 1435 | >3.6 | $43.4^{+17.7}_{-11.8}$ | $50.3^{+17.4}_{-21.0}$ | $8.1^{+0.4}_{-0.4}$ | 0.92 (108) | 5 |
| 04 | 2022-11-17 09:09:35 | 975 | >4.6 | $31.7^{+12.5}_{-6.9}$ | $45.8^{+10.1}_{-17.3}$ | $6.7^{+0.5}_{-0.5}$ | 1.30 (79) | 5 |
| 05 | 2022-12-25 05:05:36 | 1484 | >0.9 | $3.6^{+9.5}_{-3.0}$ | <12.3 | $1.8^{+0.2}_{-0.2}$ | 1.26 (24) | ... |

**Notes.** All errors represent the 90% uncertainties.
[a] The serial number indicating which joint fitting the spectrum was used for.

**Table A4**
Results of Balmer Lines Measurements from First Lijiang 2.4 m Telescope Observation

| | Flux[a] ($10^{-12}$ erg cm$^{-2}$ s$^{-1}$) | FWHM[b] (km s$^{-1}$) | $\Delta V$[c] (km s$^{-1}$) | $V_{max}$[d] (km s$^{-1}$) |
|---|---|---|---|---|
| H$\alpha$ | 39.0 ± 0.4 | 169 ± 1 | 33.1 ± 0.6 | 180 |
| H$\beta$ | 6.3 ± 0.1 | 101 ± 1 | 14.3 ± 0.5 | 105 |
| H$\gamma$ | 3.7 ± 0.1 | 122 ± 2 | 11.7 ± 0.9 | 111 |
| H$\delta$ | 3.1 ± 0.1 | 164 ± 4 | 12.8 ± 1.6 | 125 |

**Notes.** All errors represent the 90% uncertainties.
[a] The measured line flux.
[b] The measured line width after a correction of instrumental profile.
[c] The projected velocity of line center.
[d] The maximum projected velocity as the wavelength where the residual profile lies 1$\sigma$ above the continuum. The instrumental profile is corrected.

# Appendix B
# Detailed Calculations About Flaring Loop

If the flare occurred inside closed loop structures, the cooling timescale should increase with the size of the loops. As B. J. Wargelin et al. (2008) summarized, when a flare is dominated by conductive losses, its cooling timescale can be expressed as

$$\tau_C = \frac{4 \times 10^{-10} n_e L^2}{T^{5/2}}, \tag{B1}$$

where $n_e$ is the electron density, $L$ is the half-length of the flaring loop, and $T$ is the temperature of the flaring plasma. $n_e$ is related to EM by $EM = n_e n_H V \sim 0.85 n_e^2 V$ for cosmic abundances. When a flare is dominated by radiative losses, the cooling timescale is given by

$$\tau_R = \frac{3kT}{n_e \Lambda(T)}. \tag{B2}$$

Here, $\Lambda(T)$ is the cooling function of flaring plasma, whose specific form is subject to the temperature $T$. For $T \geqslant 20$ MK, $\Lambda(T) \approx 10^{-24.66} T^{1/4}$ (M. Güdel 2004).

During the fast decay phase of MF, due to the lack of constraints on $T_{hot}$ from LEIA data, we could only infer that the temperature near the flare peak is at least as high as the maximum $T_{hot}$ obtained by NICER data. At such a high temperature, conductive losses are likely the dominant factor. Therefore, we set $\tau_C$ in Equation (B1) to be $\tau_{d,MF1} = 109$ ks. Assuming the flaring plasma is confined in a single loop with constant cross section, the volume of the loop is given by

$$V = 2\pi \beta^2 L^3, \tag{B3}$$

where $\beta = r/L$ is the loop aspect ratio, and $r$ is the radius of the loop cross section. In addition, the maximum temperature $T_{max}$ inside the loop structure, electron density $n_e$, and the loop half-length $L$ obey the RTV scaling law (R. Rosner et al. 1978),

$$T_{max}^2 \approx 7.6 \times 10^{-7} n_e L, \tag{B4}$$

under the hypothesis of a loop in hydrostatic equilibrium with uniform heating. Using Equation (B1) and the RTV scaling law, the loop size $L$ and electron density $n_e$ can be both derived, as well as $V$ and $\beta$. Furthermore, the pressure $p$ inside the loop and the minimum magnetic field $B$ of the flaring plasma can be estimated by

$$p = 2n_e kT, \tag{B5}$$

and

$$B = \sqrt{8\pi p}. \tag{B6}$$

We remind that the loop size may be underestimated because the actual peak $kT_{hot}$ can be much higher than 7.4 keV. For comparison, these parameters are also derived using LEIA data assuming the peak $kT_{hot}$ of 10 keV and 20 keV. The results of the parameters are listed in Table B1. Note that $T_{max}$ is distinct from the maximum best-fit temperature $T_{obs}$, which is the average loop temperature. For simplicity, this difference is neglected in our calculations; however, we again emphasize that $T_{obs}$ could be much higher than the 86 MK observed and $T_{max}$ is somewhat higher than $T_{obs}$. Additionally, the best-fit temperatures used for calculations here are merely effective values that represent the underlying continuous distribution of temperatures in the plasma, which may also introduce uncertainty into the derived parameters.

To apply the RTV scaling law, a constant pressure throughout the whole loop is required, which means the size of loop $L$ should be smaller than the the pressure scale height of the stellar atmosphere with $H = kT/\mu m_p g$, where $\mu$ is the average mass coefficient of plasma particle ($\sim 1/2$ in a fully ionized plasma), $m_p$ is the proton mass, and $g$ is the gravitational acceleration at the stellar surface. So we derived $H \sim 1 \times 10^{14}$ cm, which is much larger than the

**Table B1**
Parameters of Loop Size and Flaring Plasma

|  | $L$[a] ($10^{12}$ cm) | $n_e$[b] ($10^9$ cm$^{-3}$) | $V$[c] $10^{36}$ cm$^3$ | $\beta$[d] | $p$[e] dyne cm$^{-2}$ | $B$[f] (Gauss) |
|---|---|---|---|---|---|---|
| MF1[g] | $1.9 \pm 0.2$ | $5.1 \pm 0.8$ | $14.0 \pm 4.6$ | $0.56 \pm 0.12$ | $120 \pm 21$ | $54.9 \pm 4.8$ |
| MF1[h] | 2.2 | 8.0 | 19.4 | 0.53 | 255 | 80.0 |
| MF1[i] | 3.2 | 22.5 | 2.4 | 0.11 | 1442 | 190 |
| MF2[g] | $3.8 \pm 0.2$ | $1.6 \pm 0.1$ | $61.4^{+11.5}_{-10.1}$ | ... | $27.3^{+3.9}_{-3.3}$ | $26.2^{+1.9}_{-1.6}$ |
| SF[g] | $0.7 \pm 0.1$ | $9.2^{+1.5}_{-1.4}$ | $0.5 \pm 0.1$ | ... | $253^{+56}_{-50}$ | $79.8^{+8.8}_{-7.9}$ |

**Notes.** Since the light curve of SF does not deviate from a single exponential decay, the conductive cooling phase is assumed to be skipped (F. Reale 2007). The cooling process of SF is then solely dominated by radiative loss. All errors represent the 90% uncertainties.
[a] Half-length of flaring loop(s).
[b] Electron density.
[c] The volume of flaring plasma.
[d] Loop aspect ratio.
[e] Pressure inside the flaring loop(s).
[f] The magnetic field strength required to confine the flaring plasma.
[g] Estimations based on NICER data.
[h] Estimations based on LEIA data, $kT_{hot,peak} = 10$ keV is assumed.
[i] Estimations based on LEIA data, $kT_{hot,peak} = 20$ keV is assumed.

$L \sim 1.9 \times 10^{12}$ cm estimated above, confirming the applicability of the RTV scaling law here.

During the cooling stage of the flaring plasma, the radiative loss gradually dominates the slow decay phase. In this phase, the initial heat pulse subsided, while a post-flare arcade built up. If we took the assumption that the loops inside the arcade all reached the same peak temperature simultaneously and had similar cooling processes, the arcade can be regarded as a single loop obeying the same cooling function $\Lambda(T)$. Then, using Equation (B2), $\tau_R = \tau_{d,MF2} = 869$ ks, $T_{hot} \approx 64$ MK ($kT_{hot} = 5.5$ keV, from the first NICER observation after $t_{p,MF2}$, observation ID = 5203530104), the electron density can be derived. Provided that the plasma is confined in a hemisphere with a radius $l = 2L/\pi$, the volume can be expressed by

$$V = \frac{32}{3\pi^2} L^3. \quad (B7)$$

The derived parameters are also displayed in Table B1.

In addition, the loop size can also be constrained by the rise timescale of the flare. During the rise phase, the high-energy electrons generated by magnetic reconnection zoomed along the loops and bombarded the footpoints continuously, violently heating the chromospheric material, which evaporated to fill the loop afterward and produced soft X-ray radiation. It is well known that the timescale of this filling process should not exceed $L/c_s$. Here, the sound speed of the plasma is $c_s = \sqrt{kT/\mu m_p}$. Using the rise timescale of the MF $\tau_{r,MF1} = 112$ ks, an upper limit of the half-length of the initial single loop is suggested to be $\sim 1.3 \times 10^{13}$ cm.

Note that the cooling process described here differs from the quasi-static cooling process (G. H. J. van den Oord & R. Mewe 1989), which requires a constant ratio between the radiative and conductive cooling timescales and the relevant flaring parameters (either $kT_{hot}$ or $EM_{hot}$) decaying exponentially. It also differs from the cooling process with sustained heating proposed by F. Reale et al. (1997), which may represent a more universal feature of stellar flares. We plan to explore these models further in our future work.

## ORCID iDs

Xuan Mao ● https://orcid.org/0009-0009-8313-1842
He-Yang Liu ● https://orcid.org/0000-0002-2412-5751
Song Wang ● https://orcid.org/0000-0003-3116-5038
Dongyue Li ● https://orcid.org/0000-0002-4562-7179
Fabio Favata ● https://orcid.org/0009-0001-8144-2526
Jujia Zhang ● https://orcid.org/0000-0002-8296-2590
Xinlin Zhao ● https://orcid.org/0009-0005-5459-7433
Jing Wang ● https://orcid.org/0000-0002-6880-4481
Mingjun Liu ● https://orcid.org/0009-0009-8982-2361
Alberto J. Castro-Tirado ● https://orcid.org/0000-0003-2999-3563
Licai Deng ● https://orcid.org/0000-0001-9073-9914
Xu Ding ● https://orcid.org/0000-0002-2427-161X
Kaifan Ji ● https://orcid.org/0000-0001-8950-3875
Chichuan Jin ● https://orcid.org/0000-0002-2006-1615
Yajuan Lei ● https://orcid.org/0000-0001-9510-3181
Huali Li ● https://orcid.org/0000-0002-7457-4192
Jun Lin ● https://orcid.org/0000-0002-3326-5860
Shuai Liu ● https://orcid.org/0000-0001-5193-1727
Jianrong Shi ● https://orcid.org/0000-0002-0349-7839
Jianguo Wang ● https://orcid.org/0000-0003-4156-3793
Jingxiu Wang ● https://orcid.org/0000-0003-2544-9544
Dingrong Xiong ● https://orcid.org/0000-0002-6809-9575
Guiping Zhou ● https://orcid.org/0000-0001-8228-565X

## References

Anders, F., Khalatyan, A., Chiappini, C., et al. 2019, A&A, 628, A94
Arnaud, K. A. 1996, in ASP Conf. Ser. 101, Astronomical Data Analysis Software and Systems V, ed. G. H. Jacoby & J. Barnes (San Francisco, CA: ASP), 17
Aschwanden, M. J., & Alexander, D. 2001, SoPh, 204, 91
Astropy Collaboration, Robitaille, T. P., Tollerud, E. J., et al. 2013, A&A, 558, A33
Bailer-Jones, C. A. L., Rybizki, J., Fouesneau, M., Demleitner, M., & Andrae, R. 2021, AJ, 161, 147
Benz, A. 2002, Plasma Astrophysics, Vol. 279 (2nd ed.; Dordrecht: Kluwer)
Benz, A. O., & Güdel, M. 2010, ARA&A, 48, 241
Boller, T., Freyberg, M. J., Trümper, J., et al. 2016, A&A, 588, A103
Candelaresi, S., Hillier, A., Maehara, H., Brandenburg, A., & Shibata, K. 2014, ApJ, 792, 67
Chen, H. C., Tian, H., Li, H., et al. 2022, ApJ, 933, 92
Cheng, H. Q., Ling, Z. X., Zhang, C., et al. 2024, ExA, 57, 10
Davenport, J. R. A. 2016, ApJ, 829, 23
Drake, A. J. 2006, AJ, 131, 1044
Drake, A. J., Graham, M. J., Djorgovski, S. G., et al. 2014, ApJS, 213, 9
Emslie, A. G., Dennis, B. R., Shih, A. Y., et al. 2012, ApJ, 759, 71
Endl, M., Strassmeier, K. G., & Kurster, M. 1997, A&A, 328, 565
Favata, F., & Schmitt, J. H. M. M. 1999, A&A, 350, 900
Feldman, U., Laming, J. M., & Doschek, G. A. 1995, ApJL, 451, L79
Franciosini, E., Pallavicini, R., & Tagliaferri, G. 2001, A&A, 375, 196
Getman, K. V., & Feigelson, E. D. 2021, ApJ, 916, 32
Getman, K. V., Feigelson, E. D., Broos, P. S., Micela, G., & Garmire, G. P. 2008, ApJ, 688, 418
Getman, K. V., Feigelson, E. D., & Garmire, G. P. 2021, ApJ, 920, 154
Güdel, M. 2004, A&ARv, 12, 71
Günther, H. M., Pasham, D., Binks, A., et al. 2024, ApJ, 977, 6
Günther, M. N., Zhan, Z., Seager, S., et al. 2020, AJ, 159, 60
Hawley, S. L., Davenport, J. R. A., Kowalski, A. F., et al. 2014, ApJ, 797, 121
Karmakar, S., Naik, S., Pandey, J. C., & Savanov, I. S. 2023, MNRAS, 518, 900
Kashapova, L. K., Broomhall, A.-M., Larionova, A. I., Kupriyanova, E. G., & Motyk, I. D. 2021, MNRAS, 502, 3922
Kiraga, M. 2012, AcA, 62, 67
Li, D. Y., Ling, Z. X., Liu, Y., et al. 2022, ATel, 15754, 1
Lin, J. 2002, ChJAA, 2, 539
Lin, J., & Forbes, T. G. 2000, JGR, 105, 2375
Ling, Z. X., Liu, Y., Zhang, C., et al. 2022, ATel, 15748, 1
Ling, Z. X., Sun, X. J., Zhang, C., et al. 2023, RAA, 23, 095007
Lyke, B. W., Higley, A. N., McLane, J. N., et al. 2020, ApJS, 250, 8
Maehara, H., Shibayama, T., Notsu, S., et al. 2012, Natur, 485, 478
Martìnez, C. I., Mauas, P. J. D., & Buccino, A. P. 2022, MNRAS, 512, 4835
Merloni, A., Lamer, G., Liu, T., et al. 2024, A&A, 682, A34
Notsu, Y., Maehara, H., Honda, S., et al. 2019, AAS Meeting, 234, 122.02
Okamoto, S., Notsu, Y., Maehara, H., et al. 2021, ApJ, 906, 72
Osten, R. A., Kowalski, A., Drake, S. A., et al. 2016, ApJ, 832, 174
Pallavicini, R., Golub, L., Rosner, R., et al. 1981, ApJ, 248, 279
Pandey, J. C., & Singh, K. P. 2012, MNRAS, 419, 1219
Pasham, D., Hamaguchi, K., Miller, J. M., et al. 2022, ATel, 15755, 1
Predehl, P., Andritschke, R., Arefiev, V., et al. 2021, A&A, 647, A1
Pye, J. P., Rosen, S., Fyfe, D., & Schröder, A. C. 2015, A&A, 581, A28
Reale, F. 2007, A&A, 471, 271
Reale, F., Betta, R., Peres, G., Serio, S., & McTiernan, J. 1997, A&A, 325, 782
Reale, F., Güdel, M., Peres, G., & Audard, M. 2004, A&A, 416, 733
Rosner, R., Tucker, W. H., & Vaiana, G. S. 1978, ApJ, 220, 643
Sasaki, R., Tsuboi, Y., Iwakiri, W., et al. 2021, ApJ, 910, 25
Shibata, K., & Magara, T. 2011, LRSP, 8, 6
Shibata, K., & Yokoyama, T. 1999, ApJL, 526, L49
Shibayama, T., Maehara, H., Notsu, S., et al. 2013, ApJS, 209, 5
Shimizu, T. 1995, PASJ, 47, 251
Smith, R. K., Brickhouse, N. S., Liedahl, D. A., & Raymond, J. C. 2001, ApJL, 556, L91
Tody, D. 1986, Proc. SPIE, 627, 733
Tody, D. 1993, in ASP Conf. Ser. 52, Astronomical Data Analysis Software and Systems II, ed. R. J. Hanisch, R. J. V. Brissenden, & J. Barnes (San Francisco, CA: ASP), 173
Tsuboi, Y., Yamazaki, K., Sugawara, Y., et al. 2016, PASJ, 68, 90
Tsuru, T., Makishima, K., Ohashi, T., et al. 1989, PASJ, 41, 679

van den Oord, G. H. J., & Mewe, R. 1989, A&A, 213, 245
Walkowicz, L. M., Basri, G., Batalha, N., et al. 2011, AJ, 141, 50
Wang, C. J., Bai, J. M., Fan, Y. F., et al. 2019, RAA, 19, 149
Wang, J., Xin, L. P., Li, H. L., et al. 2021, ApJ, 916, 92
Wargelin, B. J., Kashyap, V. L., Drake, J. J., Garcìa-Alvarez, D., & Ratzlaff, P. W. 2008, ApJ, 676, 610
Wei, J., Cordier, B., Antier, S., et al. 2016, arXiv:1610.06892
Wright, N. J., Drake, J. J., Mamajek, E. E., & Henry, G. W. 2011, ApJ, 743, 48
Xin, L. P., Li, H. L., Wang, J., et al. 2024, MNRAS, 527, 2232
Xiong, D. R., Bai, J. M., Fan, Y. F., et al. 2022, ATel, 15775, 1
Yuan, W., Zhang, C., Chen, Y., & Ling, Z. 2022, Handbook of X-Ray and Gamma-Ray Astrophysics (Berlin: Springer), 86
Zhang, C., Ling, Z. X., Sun, X. J., et al. 2022, ApJL, 941, L2